\documentclass{PoS}
\usepackage{subfig}
\usepackage[font={small}]{caption}

\title{The International Linear Collider \\- Physics \& Perspectives}

\ShortTitle{ILC - Physics \& Perspectives}

\author{\speaker{Naomi van der Kolk}\\
on behalf of the ILC Physics and Detector Study\\
        Max-Planck-Institute for Physics, Munich, Germany\\
        E-mail: \email{naomi.van.der.kolk@cern.ch}}

\abstract{With the discovery of a Higgs boson at LHC, all particles of the Standard Model seem to have been observed experimentally, yet many questions are left unanswered. The discovery has intensified the planning for future high-energy colliders, which aim to probe the Standard Model and the mechanism of electroweak symmetry breaking with higher precision and to extend and complement the search for new particles currently under way at the LHC. The most mature option for such a future facility is the International Linear Collider ILC, an electron-positron collider with a centre-of-mass energy of 500\,GeV, and the potential for upgrades into the TeV region. The ILC will fully explore the Higgs sector, including model-independent coupling and width measurements, direct measurements of the coupling to the top quark and the Higgs self-coupling, enable precision measurements of top quark properties and couplings as well as other electroweak precision measurements and provide extensive discovery potential for new physics complementary to the capabilities of hadron colliders. This paper will give an overview of the physics case of the ILC, put in context of the running scenario covering different centre-of-mass energies, and discuss the current status and perspectives of this global facility.}

\FullConference{XXIV International Workshop on Deep-Inelastic Scattering and Related Subjects\\
		11-15 April, 2016\\
		DESY Hamburg, Germany}

\begin{document}

\section{Introduction}
\vspace{-2mm}
The standard model has been very successful in describing nearly all experimental observations in particle physics. 
With the discovery of a Higgs boson at the LHC the Standard Model could be complete and could, in principle, be correct up to energies many orders higher than what is currently available.
However, the validity of the Standard Model is limited, as there remain phenomena not described by the standard model;
e.g. the nature of the Higgs field, cosmic dark matter, and the matter anti-matter asymmetry.
There is, however, no indication as to how it breaks down and at what energy. This is one of the most pressing questions in particle physics today. 
Models that try to improve upon the Standard Model all predict some sort of New Physics in the form of new particles and interactions. 
Besides looking for New Physics directly at higher energy scales or new regions of phase space, 
precision measurements of electroweak processes can provide answers to the open questions. 
By measuring very precisely the properties and interactions of the Higgs boson we can start to understand the mechanism of electroweak symmetry breaking and the nature of the Higgs potential. 
The top quark is expected to have the strongest coupling to the Higgs sector and as such its interactions and couplings are very sensitive to the presence of New Physics. 
Any electroweak process will be modified by the presence of new particles or interactions and precision measurements will show these deviations.
High precision measurements of electroweak processes can be performed at electron-positron colliders. 
These colliders offer a well defined initial state and a clean and fully reconstructable final state without background from strong interactions.
They offer a democratic environment, because the e$^+$e$^-$ annihilation produces particle pairs of (almost) all species at similar rates. 
One of the strong features of an e$^+$e$^-$ collider is beam polarisation, which gives direct experimental access to the chiral structure of produced particles.
Additionally, electroweak interactions can be calculated to a high degree of precision, making the backgrounds straightforward to compute.
The most mature proposal of such a e$^+$e$^-$ collider is the International Linear Collider.
\vspace{-2mm}

\section{International Linear Collider}
\vspace{-2mm}
The International Linear Collider (ILC) is foreseen as a 34 km long accelerator with superconducting RF structures with an acceleration gradient of 31.5 MV/m. Its design centre of mass energy is 500\,GeV, but it will be operated at other energies in order to perform precision measurements of specific processes. 
It will be possible to extend the machine and reach an energy of 1\,TeV.
At the single interaction point two detectors are foreseen to take data one after the other in a push-pull configuration.
The ILC Technical Design Report~\cite{2013_TDR3} was published in 2013 and the accelerating technology is already being used for the construction of the European Free Electron Laser (XFEL).
The Japanese government is considering to host the ILC and a candidate location has been identified in the north of Japan. A final decision is expected in the coming years.

While the actual running scenario of the ILC will depend on the physics outcomes of the LHC and of the results of the ILC itself, some baseline running scenarios have been prepared~\cite{2015_Barklow}.  
The preferred scenario, H20, is illustrated in Figure~\ref{fig:H20}. 
This program is optimised around the measurements of the Higgs boson properties and guarantees its fully model-independent characterisation by combining measurements taken at 250\,GeV and at 500\,GeV. 
Additionally the study of the top quark is guaranteed by running at the pair production threshold, around 350\,GeV, and above at the design energy of 500\,GeV.

\section{Detectors at the ILC}
\vspace{-2mm}
The physics objectives of the ILC require detectors that can reconstruct the e$^+$e$^-$ collisions with a very high precision. 
Such high precision can be achieved by using reconstruction based on the Particle Flow (PFA) paradigm~\cite{2009_Thomson}, in which each individual particle is reconstructed. This results in an excellent jet energy resolution, as for each particle in the jet the most precise energy measurement is used. 
The detectors at the ILC are designed to be optimal for PFA; they have highly granular calorimeters such that nearby particle tracks can be separated and energy depositions attributed to the correct particle, they have very precise vertex and tracking detectors for track momentum and secondary vertex reconstruction and they are as hermetic as possible for a precise measurement of missing energy.
Because of the clean environment in e$^+$e$^-$ collisions the inner tracking system can be optimised for high precision vertex reconstruction.
In particular, it is possible to push the innermost radius of the vertex detector to close to 1.5 cm, which significantly improves the flavour tagging performance.
The material budget can be very low as no cooling is required because radiation hard detectors are not needed and electronics can be operated in power pulsed mode due to the bunch structure of the ILC.

The two detectors that are foreseen at the ILC, the Silicon Detector (SiD) and the International Large Detector (ILD), are both optimised for PFA while using complementary technologies and designs. Their Detailed Baseline Design report~\cite{2013_TDR4} was published in 2013. 
The main difference between the two detectors is that ILD uses a TPC and SiD an all silicon tracker. 
As a consequence ILD is somewhat larger and uses a smaller magnetic field strength as SiD.

\section{Physics perspectives}
\vspace{-2mm}
At the ILC high precision tests of Standard Model predictions can be done over a wide energy range. 
This paper highlights some possible measurements from the vast physics program~\cite{2013_TDR2, 2015_Moortgat, 2015_Fujii}.

One of the main fields of focus is precision Higgs physics, which can provide crucial information on the process of electroweak symmetry breaking. 
In e$^+$e$^-$ collisions there are 3 Higgs production channels: Higgsstrahlung, WW-fusion and ZZ-fusion.
The Higgsstrahlung process has its maximum cross section at 250\,GeV providing about 160,000 Higgs events for an integrated luminosity of 500\,fb$^{-1}$. At 500\,GeV, a sample of 500\,fb$^{-1}$ gives another 125,000 Higgs events, of which 60\% are from the WW-fusion process. With such samples of Higgs events the rates for Higgs production and decay for all of the major Higgs decay modes can be measured.

\begin{figure}
  {\centering
    \subfloat[\label{fig:H20}]{\includegraphics[height=0.18\textheight]{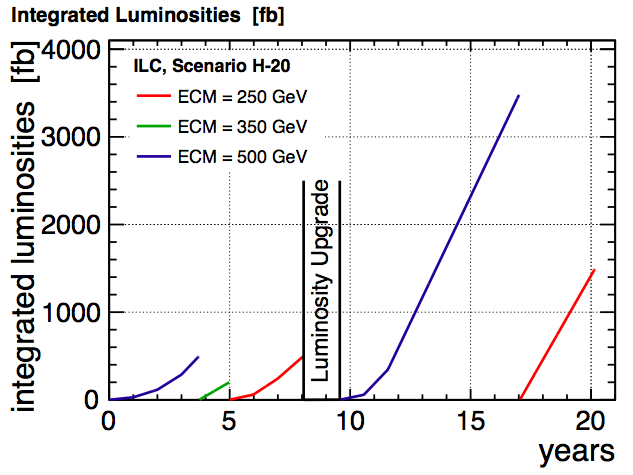}}
    $\qquad$
    \subfloat[\label{fig:Zhrecoil}]{\includegraphics[height=0.18\textheight]{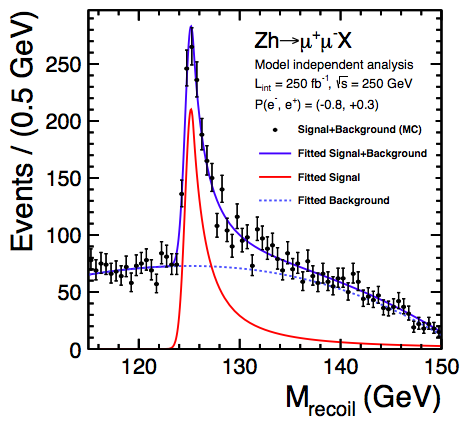}}
    \caption[]{(a) Accumulation of integrated luminosity for scenario H20~\cite{2015_Barklow}. (b) Recoil mass distribution for $e^+e^- \rightarrow  ZH$ followed by $Z \rightarrow \mu^+\mu^-$ decay for $m_H$ = 125\,GeV with 250\,fb$^{-1}$ at $\sqrt{s}$ = 250 GeV~\cite{2012_ILD_Li}.}
  }
\end{figure}

The Higgsstrahlung process, $e^+e^-\rightarrow ZH$, allows a model independent measurement of the Higgs boson properties.
Because of the well known initial state the Higgs boson can be reconstructed from its recoil against the Z boson. 
When combined with measurements in other production channels, and with the total cross section measurement, this allows the determination of the total Higgs width and the absolute normalisation of its couplings. 
Additionally Higgs decays into invisible or exotic particles can be observed in this channel.
When considering only the leptonic decay channels of the $Z$ at 250\,GeV the Higgs mass can be determined to better than 30 MeV~\cite{2013_Asner, 2012_ILD_Li}, as is illustrated in Figure~\ref{fig:Zhrecoil}.

\begin{figure}
  {\centering
    \subfloat[\label{fig:Hcouplings}]{\includegraphics[height=0.18\textheight]{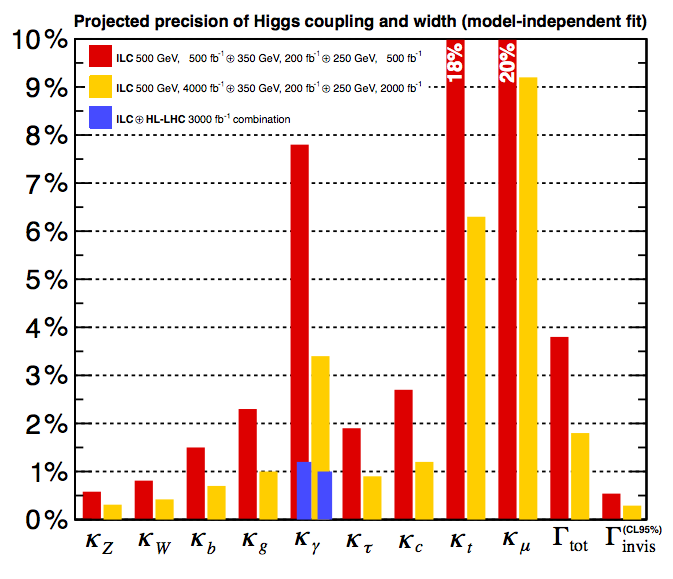}}
    $\qquad$
    \subfloat[\label{fig:Tthreshold}]{\includegraphics[height=0.18\textheight]{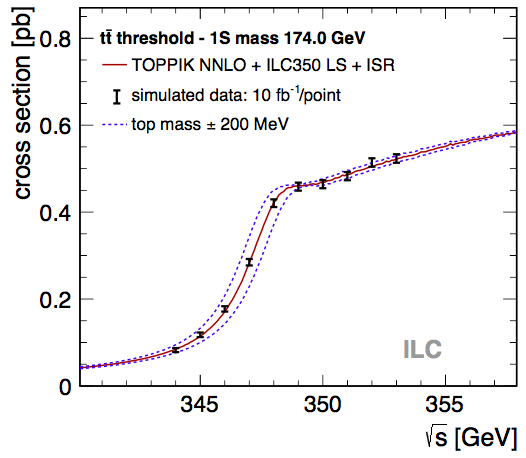}}
    \caption{(a) Relative precisions for the Higgs couplings from a model independent fit~\cite{2015_Fujii}. (b) Top quark pair production threshold corresponding in total to one year at design luminosity~\cite{2013_Seidel}.}
  }
\end{figure}

All major Higgs decay channels can be measured individually, including the decay to $c\bar{c}$ or $gg$ which are very difficult to access at hadron colliders.
By combining measurements at different centre of mass energies all couplings can be accessed in a model independent way, including the top coupling and the Higgs self coupling.
The precision with which the couplings can be measured model independently are given in Figure~\ref{fig:Hcouplings}. 
Many couplings can be measured with an uncertainty of less than 1 percent.
New particles or interactions will change these couplings with respect to the Standard Model prediction. 
Observing such deviations points to New Physics and can also provide information about the nature of the Higgs boson; composite particle or fundamental scalar particle.
Deviations from the Standard Model predictions are generally expected to be small of the order of 5\%~\cite{2015_Fujii}. 
This means that a measurement precision at the order of a percent is needed.

Another main interest is the top quark.
Being the heaviest quark it plays a key role in electroweak symmetry breaking and its mass is an important parameter in the Standard Model.
The top quark can ideally be studied at its pair production threshold around 350\,GeV. 
An scan around this energy of the production cross section, as illustrated in Figure~\ref{fig:Tthreshold}, gives a very precise and theoretically clean measurement of the top mass~\cite{2013_Seidel, 2013_Horiguchi}. 
Additionally, the shape of the production cross section depends on the top width, Yukawa coupling and the value of $\alpha_{s}$.
From this measurement a final precision on the top mass of less than 100 MeV is expected based on current studies. 
Depending on how well the experimental errors are controlled, the main contribution could come from the theoretical uncertainty in the QCD renormalisation scale.
A recent study based on NNNLO QCD calculations shows that when these scale variations are taken into account in the mass determination the resulting uncertainty is ~50 MeV~\cite{2016_Simon}. 
At energies above the production threshold all decay modes of the top quark can be measured with high resolution and very low background levels due to the clean experimental environment.
The weak and electromagnetic couplings can be measured in detail.
Beam polarisation offers extra sensitivity, because the weak couplings to W or Z depend on polarisation and thus the couplings to each polarisation state can be measured individually.
The couplings of the top quark can all be fully constrained by combining the measurements of the total production cross section, the forward-backward asymmetry and the helicity angle for two polarisation states. 
Precisions on the couplings of less than 2\% can be reached~\cite{2010_Devetak, 2013_Amjad, 2014_Richard, 2015_Amjad}.
The couplings of the top quark are affected by the existence of New Physics and expected deviations from the Standard Model predictions can be as large as 20\%~\cite{2014_Richard, 2015_Amjad}.

Searches for New Physics can be done directly or via precision measurements of Standard Model processes, particularly in the Higgs and top quark interactions.
Direct searches for the existence of new particles are possible at the ILC up to a mass of almost half the centre of mass energy. 
The ILC is complementary to the LHC in these searches as lepton colliders can extend into regions of phase space not covered by hadron colliders.
If a new particle is found or suggested the ILC can run an energy scan around the production threshold for a precise mass measurement of this new particle~\cite{2013_Baer}. 
Its quantum numbers are accessible though the pair production rates at different beam polarisations.
One area of interest is the search for Dark Matter (DM). 
Many models require the simultaneous annihilation of charged and neutral DM particles. 
The necessary small mass difference (less than 20\,GeV) between them results in soft decay particles which can be detected at the ILC. One example is stau pair production, where the stau decays into a dark matter particle and emits a soft tau lepton which is observed in its decay into low energy pions~\cite{2013_Berggren2}. 
Another way in which Dark Matter particles can be studied is via initial state radiation (ISR) photons recoiling against invisible particles~\cite{2013_Chae}.
Similar challenges are offered by the possible supersymmetric partners of the Higgs; Higgsinos. 
Naturalness requires these Higgsinos to be relatively light and almost degenerate and small mass differences between the charged and neutral states of the order of 10\,GeV are expected, resulting in soft decay products.
Higgsinos can therefore be measured from missing mass of the recoiling system against an ISR photon~\cite{2013_Berggren}.
If additional Higgs bosons exist, they can be found at the ILC if they are sufficiently light. 
They are then produced by radiating of a top or bottom quark and can be detected via their dominant $b\bar{b}$ or $\tau^+\tau^-$ decays~\cite{2005_Kilian}.
Two-fermion processes at the ILC allow to search for new gauge bosons such as the $Z'$.
If the $Z'$ exists its presence and interactions will perturb the Standard Model cross sections and at the ILC all its couplings can be determined with high precision.
Constraints on quark/lepton compositeness, that would perturb two-fermion processes, can be expected to improve by an order of magnitude~\cite{2013_TDR2}.


\section{Conclusion}
\vspace{-2mm}
This paper has shown that the ILC will be an ideal machine for precision electroweak measurements; precision studies of the Higgs boson and top quark allow for finding new particles and interactions and for finding answers to the nature of electroweak symmetry breaking.
However, ILC is also a discovery machine; direct searches for new particles are possible and complement searches at the LHC in regions that are inaccessible or particularly difficult at hadron colliders.
The technology for the accelerator is mature and the ILC is ready to be constructed. 
There is a strong interest in Japan to host this facility.

\bibliographystyle{pos}
\bibliography{proceedings}



\end{document}